# A study of the accuracy of Mass-Radius relationships for silicate-rich and ice-rich planets up to 100 Earth masses.

**Authors:** Grasset O. [1], Schneider J. [2], Sotin C. [3]

**Address :**

[1]. Laboratoire de Planétologie et Géodynamique, UMR-CNRS 6112, Université de Nantes, France.

[2]. Laboratoire Univers et Théories, UMR-CNRS 8102, Observatoire de Paris , France

[3]. Jet Propulsion Laboratory, California Technology Institute, Pasadena, CA.

# Abstract

A mass-radius relationship is proposed for solid planets and solid cores ranging from 1 to 100 Earth-mass planets. It relies on the assumption that solid spheres are composed of iron and silicates, around which a variable amount of water is added. The M-R law has been set up assuming that the planetary composition is similar to the averaged composition for silicates and iron obtained from the major elements ratio of 94 stars hosting exoplanets.

Except on Earth for which a tremendous amount of data is available, the composition of silicate mantles and metallic cores cannot be constrained. Similarly, thermal profiles are poorly known. In this work, the effect of compositional parameters and thermal profiles on radii estimates is quantified. It will be demonstrated that uncertainties related to composition and temperature are of second order compared to the effect of the water amount.

The Super-Earths family includes four classes of planets: iron-rich, silicate-rich, water-rich, or with a thick atmosphere. For a given mass, the planetary radius increases significantly from the iron-rich to the atmospheric-rich planet. Even if some overlaps are likely, M-R measurements could be accurate enough to ascertain the discovery of an earth-like planet .The present work describes how the amount of water can be assessed from M-R measurements. Such an estimate depends on several assumptions including *i)* the accuracy of the internal structure model and *ii)* the accuracy of mass and radius measurements. It is shown that if the mass and the radius are perfectly known, the standard deviation on the amount of water is about 4.5 %. This value increases rapidly with the radius uncertainty but does not strongly depend on the mass uncertainty.

# I. Introduction

Observing other planetary systems opens the perspective of discovering new types of planets which are unknown in our Solar System. The recent discoveries of exoplanets with masses below 20 Earth masses (Santos et al. 2004; McArthur et al. 2004; Rivera et al. 2005; Beaulieu et al. 2006; Lovis et al., 2006; Udry et al. 2007; Bennett et al., 2008; Mayor, 2008) confirms the existence of « Super-Earths », a word that has been used by exoplanetologists for a few years (e.g. Melnick et al. 2001). In agreement with Seager et al. (2007), the Super-Earths family considered in this work includes not only terrestrial planets such as Mars, Venus and Earth (Sotin et al, 2007), water-rich planets which are similar to giant icy moons (Léger et al., 2004), and iron-rich planets similar to Mercury (Valencia et al., 2006), but also "mini-Neptunes".

The abundance of low mass planets is not yet known. So far, over the 304 exoplanets which have been discovered, more than 90 % are giant planets, and only 26 are below 30 Earth masses. But this observed distribution, which is biased by observational limitations, is evolving very rapidly. First, low mass planets are preferentially in multi-planet systems (10 out of the 13 planets with masses below 20 $M_{Earth}$ are in multi-planet systems). It indicates that planetary systems have generally several low mass planets. Second, preliminary results from the HARPS spectrograph suggest that more than 30% of stars host Super-Earths (Mayor, 2008). Finally, planets around 5 Earth masses can now be detected as shown by the discoveries of Gliese 581c (Udry et al. 2007), OGLE-2005-BLG-390Lb (Beaulieu et al., 2006), MOA-2007-BLG-192 (Bennett, 2008), and HD40307a-c (Mayor, 2008).

Planets can be either gaseous or solid. Some overlap should exist between these categories and we argue that large planets, several tens of Earth masses, are not necessarily gaseous planets. The presence (or absence) of an atmosphere depends indeed on the interplay between three factors: primordial gas accretion (Benz, 2006), outgassing (Musselwhite and Drake, 2000), and atmosphere erosion by the stellar wind (Erkaev et al., 2007). Zuckerman et al (1995) have shown that stellar winds may lead to the inhibition of giant planet formation by rapid gas depletion. Since the structure of a giant planet consists of a solid core surrounded by a large primordial atmosphere of $H_2$ and He, a strong depletion can result in a very massive naked core which bears many similarities with solid planets discussed in the present paper. Recently, Baraffe et al. (2008) have discovered such a 200 Earth-mass planet within HAT-2 b.

Previous work by Zapolski and Salpeter (1969) and Fortney et al. (2007) have proposed mass-radius relationships in order to distinguish between gaseous and solid planets over a very large mass range. Several models have also been developed for solid bodies of moderate masses (Léger et al., 2004; Valencia et al., 2006; Sotin et al., 2007; Seager et al, 2007; Adams et al., 2008). These studies use robust equations of state (EOS's) which allow the calculation of densities over a very large range of pressure and temperature. Although different assumptions for the planet composition are used, the results are generally in good agreement. A power-law relation of the form $R/R_0=a.(M/M_0)^b$ can be found for each family with a power exponent which decreases from 0.3 for low-mass exoplanets down to 0.2 for large-mass exoplanets (above 70 Earth masses), in agreement with the pioneering work of Zapolski and Salpeter (1969). The parameter (a) strongly depends on the planet composition, but its dependence is difficult to quantify because the models differ in many ways : silicate composition, thermal profile, metallic core mass fraction, choice of equation of state, and range of investigated masses. Both the thermal structure and the precise composition will not be known in the near future since these parameters are still debated for the Earth. In this work, the effects of these different inputs are quantified in order to investigate the possibility for M-R relationships to distinguish between water-poor (like Earth) and water-rich (Ocean-planets) solid planets. It illustrates how accurate will be the determination of the composition of a solid planet in the near future.

This study focuses on silicate and water rich planets. By analogy with the solar system, a metallic iron core is envisaged if the amount of iron is sufficient. One important aspect of these planets is that they are potentially « habitable », in the sense of bearing a sufficient amount of liquid water at the surface (if permitted by the planetary surface temperature). It is anticipated that terrestrial concepts such as oceans and continents, volcanic and tectonic activity and even vegetal cover (« forests ») are applicable to these planets. Planets having a mass lower than one Earth' mass, have probably lost their liquid (and vapour) water and are thus not habitable, loosing their exobiological interest (Huang 1960). Interestingly, Earth may lie at the bottom of an « habitability sequence » since on average, habitable exoplanets are probably larger than the Earth. In addition, planets larger and more massive than the Earth are easier to detect, whatever the detection method: radial velocity, astrometry, transits, microlensing or direct detection (Selsis et al., 2007; Schneider 1999). These are the two reasons leading us to investigate Super-Earths rather than « Mini-Earths ».

It is assumed that the structure of a super-Earth is similar to that of the Earth and other planets and icy satellites of the solar system: a core composed of an iron-rich alloy, overlaid by a silicate mantle, and a water-rich outer layer. The bulk composition is then fixed by the amount of water, and the atomic ratios Fe/Si and Mg/Si (see section II). Stars which host exoplanets, have Fe/Si and Mg/Si ratios which can differ significantly from solar ones (Beirao et al., 2005; Gilli et al., 2006). It is shown that the present model is compatible with all these compositions, which means that the presence of silicate-rich planets with or without water is very likely around these stars. In addition, the variations of Fe/Si and Mg/Si ratios are investigated and their implications on the M-R relationships are quantified.

Section II describes the chemical composition of planetary interiors and the modelling. A reference model assumes specific composition and physical parameters (Table 1). In section III, M-R relationships are inferred using the reference model. Section III also describes the influence of compositional and temperature variations on the scaling law. In a last part, results are compared with those of previous studies. In addition, it is shown that future exploration programs may allow for a rough characterization of the solid part of discovered exoplanets if the atmospheric contribution can be removed.

## II. Internal structure of super-Earths

*II.1. Composition of planets*

Massive planets are assumed to be fully differentiated and composed of an iron rich core, a thick silicate layer, and possibly a water-rich layer which draws the transition from Earth-like to icy-rich planets (Figure 1). Both silicate and water layers are split into two sublayers in order to take into account both high and low pressure mineralogy. This slight complexity avoids an underestimate of the radii. It is not fully required for very massive planets where low pressure phase layers are very thin but is important for planets below 10 Earth masses ($M_E$). As an example, the upper silicate layer composed of low-pressure phases on Earth occupies 35 % of the total volume of the mantle.

In the solar system, silicate mantles and iron cores are composed of mineral phases which mainly combine eight elements: Si ($3.548\ 10^{-5}$), Mg ($3.802\ 10^{-5}$), Fe ($3.467\ 10^{-5}$), O ($8.51\ 10^{-4}$), Ca ($2.291\ 10^{-6}$), Al ($2.951\ 10^{-6}$), Ni ($1.778\ 10^{-6}$), and S ($1.862\ 10^{-5}$). Numbers indicated in parenthesis are the atomic abundance of each species relative to hydrogen in the solar atmosphere (Cox et al., 1999). The elementary abundance of a planet is supposed to be the stellar abundance, as it is the case for the meteorites when their composition is compared to that of the Sun's photosphere (Cox et al., 1999; Lodders et al., 2003). Sotin et al (2007) have shown that a planet can be described with great accuracy using only the four elements (Si, Mg, Fe, O) and adding (Ca, Al, Ni, S) to their closest major element (Ni behaves like Fe, most of the sulphur is present in the iron core, Al is equally divided between Mg and Si in order to account for charge conservation, and Ca is added to Mg). In addition, hydrogen must be added for introducing the water layer above the silicates. The (Fe/Mg) ratio is supposed constant throughout the mantle (Sotin et al., 2007) and identical in each phase. Its value is fixed by the $Mg_\#$ parameter which is defined by:

$$Mg_\# = \left( \frac{Mg}{Mg + Fe} \right)_{silicates} \qquad (1)$$

Using the simplified composition proposed by Sotin et al. (2007), planetary composition is fixed with only four parameters : [Fe/Si], [Mg/Si], $Mg_{\#}$, and the amount of water in the planet. Solar values of [Fe/Si] and [Mg/Si] are 0.977 (0.986) and 1.072 (1.131) respectively, with the values in parenthesis being the ratio once Ca, Al, Ni, and S are added to their closest major elements. Throughout the paper, these corrections, which are relevant for silicate material, are imposed.

Distribution of elements within each layer is not straightforward. The water compound must be mainly put in the two upper layers which are considered as low and high pressure icy mantles (Figure 1). But a small amount of water could be incorporated in hydrated silicates at moderate pressures. On Earth, the amount of water which is trapped in the upper mantle is very small and still debated (Manning, 2006). Tajika (1998) suggests a total amount roughly equal to the water abundance in the oceans while Karato (2004) argues that a large amount of water, up to 10 times the ocean content, can be present in Earth's mantle. Even with this very high estimate, the water which is trapped on Earth's mantle represents less than 0.23 % of the planetary mass. Such a small relative amount leads us to neglect the amount of water present in the silicates.

Distribution of chemical compounds (Si, Mg, Fe, O) between the silicate mantle and the iron core is not fully known and depends on the processes involved in the differentiation of the core. Whereas magnesium can only be incorporated in silicates, the other (Fe, O, Si) elements can be located either in silicates or in the metallic core, with Fe going preferentially in the core and O and Si in the silicates. Following Sotin et al. (2007), we propose a simplified mineralogy for the silicate mantle. The lower mantle is composed of perovskite ($[Mg,Fe]SiO_3$) and magnesowüstite ($[Mg,Fe]O$). The upper mantle is modelled with two phases: olivine ($[Mg,Fe]_2SiO_4$) and pyroxene ($[Mg,Fe]_2Si_2O_6$). The whole mantle is assumed chemically homogeneous and the limit between upper and lower mantle is only due to the pressure controlled transformation of olivine to perovskite at 23 GPa.

Although iron is the major component of the metallic core, a large amount can be trapped in the mantle (for example, $Mg_{\#} \sim 0.7$ on Mars). Si and O are preferentially in the silicate matrix but can also mix with iron in the metallic core. Silicium has been proposed first for explaining the low density of the Earth's core (Birch, 1964) and several studies suggest that more than 7 %wt of Si is trapped in the core (e.g. Allègre et al., 1995; Javoy, 1995). Oxygen is not soluble with iron at ambient condition, but at high pressure and temperature it can form iron oxide FeO. Ohtani et al. (1984) have shown that 8.7 %wt of oxygen can mix with iron at 20 GPa and 2500 °C. Sulphur can be incorporated in the core as FeS. Sulphur abundance cannot exceed 3 %wt in the Earth's core (Dreibus and Palme, 1996; Allègre et al., 1995; Javoy, 1995), but much larger values (around 14 %wt) are expected on Mars (Sohl and Spohn; 1997). Since it is not possible to constrain the amount of volatiles which are trapped in the metallic core of exoplanets, the model described in section II, assumes that the core is made of pure iron (Table 1). This assumption allows us to compare the accuracy of several equation of states under extreme conditions. The effect of volatiles on M-R relationships will be assessed in section III. It will be shown that it influences the size of the core but only slightly the size of the planet.

First, the diversity of planetary compositions is investigated by looking at the chemical abundances of 94 stars with planetary companion (Fig. 2). The data come from Beirao et al. (2005) and Gilli et al. (2006). Measured ratio are corrected by adding the contribution Ca, Al, and Ni to their closest major element. Stars have Mg/Si ratio that vary between 0.55 and 1.85 and Fe/Si ratio between 0.5 and 1.3. The population is distributed around Fe/Si=1.01 and Mg/Si = 1.34 (white dot). While Fe/Si is similar to the solar value (0.99), Mg/Si is significantly larger (1.13). It is worth noting that all these data points are within the validity domain of the planetary model proposed above. Ranges of Fe/Si and Mg/Si ratio for specific internal structure without hydrosphere are plotted depending on the weight percentage of the core and the $Mg_\#$. With the silicate mineralogy described above, the Mg/Si ratio is allowed to vary within the range [0.6 - 2.0]. Therefore, all the planets which can be derived from the known compositions of the stars are compatible with our petrological model. There is no upper limit on the Fe/Si ratio since large values can be accounted for by the relative size of the core. For example, Mercury which has a Fe/Si ratio around 8, has a relatively large core An increase in the $Mg_\#$ parameter allows for higher and broader values of Mg/Si. Distribution of Fe/Si around 1.0 is compatible with planets possessing a metallic core (20-30 % wt). For $Mg_\#$=0.8, almost 90% of the star compositions are compatible with a silicate-rich planet, while the remaining data on the right with high Mg/Si values require higher Mg# to be explained. In this paper, we assume that there is always a metallic core. Nonetheless, if both Fe/Si and Mg/Si are low, the core is not required if Mg# is very low (bottom left part of stellar compositions in Figure 2). We have chosen not to consider this extreme case because it corresponds to a very small amount of data. In addition, solutions are not unique and both low Fe/Si and Mg/Si ratio can be explained with moderate Mg# (~0.8) and small metallic cores.

As a reference (Table 1), we propose to take [Fe/Si]=1.01 and [Mg/Si] = 1.34, well centred within the proposed distribution of Gilli et al. (2006). The Mg number is fixed to 0.8 which means that silicates are mostly forsteritic and that the iron is mainly located into the core. This hypothesis is very realistic because iron is known to mix with magnesium within silicates. Its value is around 0.9 on Earth and 0.7 on Mars.

*II.2. A brief description of the model*

The model used in this paper is mainly based on the previous version described in details in Sotin et al. (2007). Input parameters are described in Table 1. For a given composition and a planetary mass, internal structure of a planet is computed using the following set of equations:

$$\begin{cases} M = 4\pi \sum_{i=1}^{5} \int_{R_i}^{R_{i+1}} r^2 \rho_i(r) dr \\ \rho_i(r) = f_{EOS}(P_i(r), T_i(r)) \\ P_i(r) = P(R_{i+1}) + \int_{r}^{R_{i+1}} \rho_i(x) g_i(x) dx \\ g_i(r) = g(R_i)\left(\frac{R_i}{r}\right)^2 + \frac{4\pi G}{r^2} \int_{R_i}^{r} x^2 \rho_i(x) dx \\ T_i(r) = T(R_{i+1}) \exp\left[\int_{r}^{R_{i+1}} \frac{\gamma}{\Phi} g_i(r) dr\right] \end{cases} \qquad (2)$$

with r the radius, $R_i$ the position of the interfaces between layers i-1 and i, $\rho$ the density, $P$ the pressure, $T$ the temperature, $g$ the acceleration of gravity, and $\gamma$ and $\Phi$ the Grüneisen and the seismic parameters, respectively (Poirier, 2000). This set of equations is solved iteratively by adapting the size of the core in order to fit the imposed Fe/Si and Mg/Si ratio. Details of computation and parameter values can be found in Sotin et al. (2007). This model has been validated using solar system bodies (including large icy moons) for both Earth-like and water-rich planets (Léger et al., 2004; Sotin et al., 2007).

Sotin et al. (2007) proposed mass-radius relationships for Earth-like and icy planets up to 10 Earth masses. The present work expands the previous work of Sotin et al. (2007) to masses up to 100 Earth mass and investigates the precision of scaling laws for future interpretations of (M-R) observations. The most important improvement is related to the equations of state (EOS) and is detailed below.

*II.3. Equations of state*

In the set of equations (2), the second line is the relation between temperature, pressure, and density, for each material. As shown in Seager et al. (2007), two different domains can be distinguished. At low pressures (< 200 GPa), different (P-T-$\rho$) relations are used: Mie-Grüneisen-Debye (MGD), Vinet, and Birch-Mürnaghan formulations. The parameters of these EOS are well constrained by experimental data. In addition, these equations partly take into account the complex chemistry of the metallic core and the silicate mantle because EOS have been constrained on a large amount of minerals (Seager et al., 2007; Sotin et al., 2007; Valencia et al., 2006, 2007a, Léger et al., 2004). Vinet and MGD formulations are quite robust when extrapolated to pressures up to 1 TPa.

At very high pressures (> 10 TPa), first principles equation of state are commonly used such as the one derived from the Thomas–Fermi–Dirac (TFD) theory, in which the atomic structure of the solid is not considered, and the *ab initio* quantum mechanical theory for specific minerals (e.g. Poirier, 2000). Salpeter and Zapolsky (1967) improved the original TFD formulation in order to take into account some effects related to the interaction between electrons. This approach has been used recently by Fortney et al. (2007) for studying M-R relationship up to 1000 $M_E$.

The problem is mainly in the range [200 GPa – 10 TPa] where no EOS is available. Although the Rankine – Hugoniot EOS well describes shock-wave experiments, the extrapolation necessary to provide isothermal EOS is difficult. Alternatively, Seager et al. (2007) extrapolate low pressure EOS up to the pressure where it crosses the TFD. For ice, a more sophisticated interpolation process is described. A third option, chosen in the present paper, is to use the ANEOS code for filling up the gap between low (P<200 GPa) and high (P>10 TPa) pressure domains. ANEOS was initially set up for shock physics studies and provides analytic equation of state for many materials (Thompson, 1990).

The density of water, $MgSiO_3$, and iron at 300K is represented as a function of pressure below 100 TPa (Figure 3). For each material, the three EOS (MGD, ANEOS, TFD) are compared. Several tests at high temperatures up to 6000 K (in the solid domain) have been conducted with the same conclusions. The grey domain represents the pressure range in which the compound can be observed for planets less massive than 100 $M_E$. TFD curves are derived from Seager et al. (2007). MGD data are issued from the previous work of Sotin et al. (2007). First, it appears that ANEOS and TFD are very close above 1 TPa except for pure iron where it is close only above 10 TPa. For pure iron, the MGD formulation can be extrapolated up to 40 TPa, which is a surprisingly high value. ANEOS provides iron densities in very good agreement with both MGD estimates at moderate pressure and TFD theory above 10 TPa. Thus, it has been chosen for describing the entire metallic cores.

In the silicate layer, the MGD extrapolation is close to both ANEOS and TFD theory up to 100 TPa. This result differs from the tests described in Seager et al. (2007) where strong differences occur above 28 TPa. This difference between the two studies is explained by the parameters which are used in the MGD formulation. Several sets of parameters can fit the experimental constraints at moderate pressures up to 200 GPa. But depending on the chosen set, extrapolation to higher pressures may differ significantly. One important feature of the MGD formulation (or equivalently the Vinet and the Birch-Mürnaghan theories) is that the chemistry of the silicates, specifically the iron amount, can be taken into account in order to calculate the bulk density of the silicate mantle. The difference is small at moderate pressure and negligible above 200 GPa (see for example figure 2 in Seager et al., 2007). Thus, the MGD formulation is justified for low-mass planets where the density of the silicate mantle, which is under moderate pressures, is influenced by its iron content. In this range which is below several tens of GPa, both our formulation and the one proposed by Seager et al. (2007) are in perfect agreement. In the transition zone between 110 GPa and 10 TPa, and above 10 TPa, the ANEOS formulation has been chosen. It is close to our MGD estimates up to 10 TPa. Above this value, ANEOS and TFD provide the same densities. Compared to previous studies, the chosen MGD formulation for silicates under low pressure is comparable with the EOS used by Valencia et al. (2006), and Seager et al. (2007), and in agreement with high pressure experiments data. In the transition zone between 200 GPa and 2 TPa and at very large pressures, density estimates of the pure forsterite (without any iron in the silicates) using ANEOS are consistent with the values proposed by both Seager et al. (2007) and Fortney et al. (2007). In the silicate mantle, the transition of perovskite into post-perovskite above 120 GPa (Shim, 2008) is not considered. Post-perovskite may be the most abundant phase within silicate mantles of Super-Earths, but due to the lack of experimental data on the equation of state, the perovskite phase has been preferred. We argue that this simplification does not change strongly M-R relationships because density measurements of post-perovskite and perovskite in the P-T conditions of the Earth mantle differ by less than 1.2 % (Murakami et al., 2004). On the other hand, post-perovskite may be able to incorporate more iron than perovskite, which could imply a slight increase of the planetary radii at large masses. But in absence of robust experimental constraints, we prefer not to investigate this effect.

For ices at pressures below 45 GPa, MGD provides an accurate estimate of the ice density (Hemley et al., 1987). In the moderate pressure range, there is a gradual transformation of ice VII to ice X and MGD, ANEOS, and TFD provide estimates of the ice density which differ significantly. In fact, ANEOS densities are close to the proposed EOS from Belonoshko and Saxena (1991), which has been set up using molecular dynamics in the super-critical domain. Above 8 TPa, ANEOS and TFD provide similar density estimates. Thus, we choose the ANEOS formulation for both the transitional domain and the very large pressure domain. In the transitional domain between 45 GPa and 8 TPa, Seager et al. (2007) prefer using a regular transition from MGD to ANEOS which is based on *ab initio* simulations, but the density estimates are actually very similar. Without any experimental data in this pressure range, it is impossible to know which approach is better than the other. But in any case, the fact that M-R scaling laws for water-rich planets are very similar in both studies (see figure 9 and section IV.2) indicates that this uncertainty on the EOS of ices in the transitional domain is not critical for setting up M-R relationships.

## III. Mass - radius relationships for super-Earths and icy worlds

*III.1 The reference case*

Planetary radii have been calculated for a range of masses and amount of water ice ($X_w$) in the reference case (Table 1 and Figure 4) . The first result is that radii are only multiplied by 3 when the mass is increased from 1 to 100 $M_E$, whereas it should be $100^{1/3} = 4.6$ if the sphere was homogeneous and incompressible. The exponent decreases as mass increases resulting in a significant flattening of the curves above 20 $M_E$. For a given mass, the transition from a dry silicate-rich planet to an icy world induces an increase of the radius between 30 % (100 $M_E$) and 40 % (1 $M_E$). All this information can be gathered into one empirical equation which links the mass to the radius using parameters which depend on the amount of water:

$$\log\left(\frac{R}{R_E}\right) = \log(\alpha) + \left(\beta + \gamma\frac{M}{M_E} + \varepsilon\left(\frac{M}{M_E}\right)^2\right)\log\left(\frac{M}{M_E}\right) \qquad (3)$$

with $M_E$ the mass of the Earth, $R_E$=6430 km the radius of a one Earth mass planet in the reference case (but not the Earth's radius which is 6371 km), α, β, γ, and ε parameters which depend on the amount of water $X_w$ according to a power 2 law:

$$\xi = \sum_{i=0}^{2} \xi_i X_w^i \qquad (4)$$

where ξ symbolizes either α, β, γ, or ε (Table 2). Equations (3-4) with parameters of Table 2 describe M-R relationships estimated from the model in the range 1-100 $M_E$ for any planet ranging from Earth-like to pure ice planets with an accuracy better than 1 % (Figure 4).

### III.2. Effect of composition parameters

Since stars with orbiting planets have different amounts of major elements, we have investigated the effect of the elementary composition on the core size and the radius of the silicate mantle. Variations of both core and silicate radii are calculated for a large range of composition distributed randomly about the reference values (Table 1). Masses of 1, 10, 50, and 100 $M_E$ have been tested and it is worth noting that results are similar, whatever the mass is. As expected, increasing the Fe/Si ratio results in a decrease of the planetary radius and an increase of the core radius (top panel in Figure 5). For a ratio Fe/Si twice larger, the radius decreases of 1.5 % only but the inner metallic core is increased by 20 %. The effect of Mg/Si is illustrated in the second plot. It is quite small and opposite to the Fe/Si ratio. For Mg/Si twice larger, the silicate mantle radius increases by 2.5 % and the inner core radius decreases by 10 %. Finally, the influence of the amount of volatiles which may be trapped within the metallic core, has been investigated. If a large amount of volatiles is incorporated within the metallic core, its density decreases and its size increases. If 40 % of volatiles are incorporated, the size of the metallic core increases by 25 % while the radius of the silicate mantle is increased by only 1 % (third panel in Figure 5). The bottom diagram represents the influence of the iron content within the silicates ($Mg_\#$). The radius of the silicate layer is almost insensitive to this parameter which strongly affects the size of the metallic core since it controls the amount of iron which is incorporated within the silicates. If the amount of iron is increased within silicates (lower $Mg_\#$), it implies logically a significant reduction of the metallic core.

The main result is that compositional effects in silicates and metallic alloys influence only slightly the global size of the dry planet (iron core + silicate layer). From left to right, the planetary radius is always within the range ± 2 % of the reference value. On the contrary, radius of the core is very sensitive to the composition of silicates and alloys. Variations as large as 10 % can be expected depending on the amount of iron and volatiles within the planet. But the most important parameter is the $Mg_\#$ which implies variations of the core size as large as 30 %.

### III.3. *Effect of temperature variations*

The effect of temperature variations is plotted in figure 6 for a 10 Earth mass planet with $X_w$= 0.01 %wt. Ninety thermal profiles have been produced randomly within a plausible range by varying both temperature drops at the interfaces and adiabatic slopes within the different layers (Table 1). The warmest thermal profile is actually not realistic since silicate mantle is almost fully liquid. Similarly, the coldest case is close to the case of complete cooling, for which no heat sources (e.g. gravitational energy, tidal heating, or radiogenic heating) remain within the planet. It may be possible, but only for small and very old bodies.

Core radius is plotted versus planetary radius at the bottom of figure 6. It can be seen that the effect of temperature profiles is very small for the two radii since the largest variations from the reference case are lower than 1.5 %. Similar tests have been conducted in the same way for different masses leading to the same conclusions.

*III.4. Determination of the water amount in a solid planet : model uncertainties*

The two previous sections demonstrate that the radius of an Earth-like planet is not strongly affected by reasonable compositional variations and temperature differences. In fact, the radius varies by less than 5 % from the smallest possible case to the largest one, whatever the mass of the planet is. If the mass and the radius of the solid part of a planet are precisely measured and if the atmospheric contribution can be removed, the amount of water can be determined. In order to estimate the accuracy of this water amount, 530 simulations have been performed. For each simulation, the different parameters are randomly fixed in the ranges indicated in the third column of Table 1. Distribution of the different parameters is plotted in figure 7. In each case, the planet is described by its mass $M_p$, its global composition ($X_w$, Fe/Si, Mg/Si, $Mg_\#$) and a randomly fixed temperature profile. The model computes the whole internal structure and the planetary radius $R_p$. Since the characteristics of the planet are not those of the reference case, the computed radius $R_p$ differs from the theoretical radius computed with equations (3-4). Alternatively, if one assumes that ($M_p$, $R_p$) are the measured mass and radius of a given planet, it is possible using equations (3-4) to estimate iteratively the amount of water of this planet $X_w^*$. Using the 530 simulations plotted in figure 7, and assuming that the ($M_p$, $R_p$) measurements are exact, the distribution of the differences $X_w$-$X_w^*$ can be obtained. It follows a gaussian distribution with an average value almost equal to 0 and a standard deviation $\sigma_x^{\text{mod}} = 0.045$ (figure 8).

Thus, we argue that the amount of water of a Super-Earth can be determined with a standard deviation of 4.5 % as long as its mass and radius are known exactly. This standard deviation is the result of the different unknowns in the internal structure in term of composition and temperature. It is a very low value, which illustrates well the fact that silicate composition and temperature profiles are of second order importance relative to the amount of water for M-R relationships.

## IV. Discussion

A description of the mass-radius relationship for super-Earths has been developed in the mass range [1 – 100 $M_{Earth}$]. The accuracy of the model in terms of compositional and thermal uncertainties has been investigated. In this section, the relevance of this study for future super-Earths exploration is discussed.

*IV.1. Theoretical mass-radius relationships*

After the pioneering work of Zapolsky and Salpeter (1969), M-R relationships for cold spheres had not risen scientific interest for several decades. But the recent discovery of low-mass exoplanets and the future possibility of discovering super-Earths have motivated the search for more precise M-R relationships. Léger et al. (2004), and Sotin et al. (2007) have proposed M-R relations for Earth-like and water-rich planets at low masses (< 10 $M_{Earth}$). Valencia et al. (2006) chose to compare Mercury-like planets with Earth-like planets. Fortney et al (2007) provided M-R relationships for masses ranging from 0.01 Earth masses to 10 Jupiter masses.

In previous studies, it has been shown that for planets less than 10 Earth masses, a simple power law relation can be found (Valencia et al., 2006; 2007b; Sotin et al., 2007). Compared to equation (3) and to the results from Seager et al. (2007), these previous works neglect the influence of the mass on the exponential factor ($\varepsilon = \gamma = 0$) because they are dealing with planets of low masses. But taking into account larger masses induces a significant decrease of the power exponent. A flattening of the M-R curves is expected at large masses because compression effects become more and more important (Zapolsky and Salpeter, 1969; Seager et al., 2007; Fortney et al., 2007). Without going to very large masses (around 1000 $M_E$) where the materials begin to behave like a Fermi gas (Fortney et al., 2007), the compression effect is actually noticeable at low masses. A possible way to take this effect into account is to use a constant exponent for the power law (1/3) and add a correction term which increases with mass (Seager et al., 2007). But we prefer the formulation proposed in equation (3) because it illustrates well how the power exponent decreases when mass increases.

*IV.2. Influence of planetary composition on M-R relationships*

Zapolski and Salpeter (1969) studied planetary M-R relations assuming homogeneous spheres. In the recent work of Seager et al. (2007), mass-radius relationships are investigated for cold planets made primarily of iron, silicates, water, and carbon compounds over a very wide range of masses. Léger et al. (2004), Valencia et al. (2006, 2007b), and Sotin et al. (2007) used complex chemistry for silicates and introduced thermal effects. The effect of large amounts of water on M-R relationships was first investigated by Léger et al. (2004). In figure 9, the different estimates of M-R relationships in the range 1-100 $M_E$ have been plotted. The main result is that all these studies are in good agreement. The slight differences can be explained by the fact that models use different temperature fields, different modal compositions for silicates, and different equations of state. In fact, figure 9 illustrates well the fact that precise compositions and/or thermal profile choices are of second order importance in M-R relations.

Nonetheless, two significant discrepancies can be noted: in the pure water case for very large masses, and in the pure silicate case at low masses. In the case of icy planets, Fortney et al. (2007) proposed an M-R relationship which is significantly above the others. They used an empirical equation of state corrected of thermal effects along the Uranus/Neptune adiabat (Guillot, 2005), while in Seager et al (2007) and in this work, a combination of low and high pressure equation of state is preferred. Seager et al. (2007) and the present work use equations of state which are not empirical, especially at low pressure where the density strongly influences the radius. Both studies are in very good agreement with each other. Although Fortney et al (2007) applied the EOS over a much wider range of masses, we suggest that their estimate of radii in the pure water case and in the [ 1 – 100 $M_{Earth}$] domain is overestimated.

In the case of silicates under 10 Earth masses, the M-R curve from Seager et al. (2007) is above the others. In particular, it is above the Earth point (1 Earth radius for 1 Earth mass). In fact, this curve is the M-R relationship for a pure forsterite planet, that is without iron. In the studies of Valencia et al. (2007b), Sotin et al. (2007) and this work, an iron core is always considered, which reduces significantly the radius of low mass planets compared to the ideal silicate case. In addition, our model includes some iron in the silicates ($Mg_\#$=0.8 in the reference case), which makes the silicates denser than pure forsterite. It is worth noting that these two points become negligible above 20 Earth masses because the results from Seager et al. (2007) are very similar to those of Fortney et al. (2007) and ours.

In figure 2, it has been shown that the compositions of the stars which host exoplanets are compatible with planets composed of an iron core, a silicate mantle and eventually an hydrosphere and an atmosphere. Silicates are made of pyroxene and olivine or eventually pyroxene and quartz for very low Mg/Si ratio. But one must consider that planets may not be necessarily composed of a mixture of iron, silicates, water and He-H gas. In particular, carbon rich planets can be envisaged in some stellar environments. To consider such new families is certainly relevant, especially from the astrobiological point of view, but it is worth noting that it will be very hard to distinguish this specific family from Super-Earths. First of all, the carbon either in its graphitic form or mixed with silicon, has density similar to a mix of silicates and water as is shown in Figure 9 using Zapolky and Salpeter data (1969) for pure carbon. If it mixes with water forming hydrates and or clathrate hydrates, densities are almost equal to water densities, at least at low pressures where these phases are known to exist. Seager et al. (2007) have indeed shown that carbon- rich planets are located within the Super-Earth domain. That is why these specific planets have not been investigated in this work.

*IV.3. How to distinguish the different families in the future?*

For a given mass, the size of a gaseous planet is roughly three times larger than a planet from the Super-Earths family. In this class of planets, the radius of a Mercury-like planet is about 80 % that of an Earth-like planet (Figure 9). Radii also strongly depend on the amount of ices. In fact, it could be possible to characterize Super-Earths composition (Mercury-like, Earth-like, water-rich, or mini-Neptune) from M-R measurements. Nonetheless, solutions are not unique. As an example, it must be noted that a silicate-rich planet surrounded by a very thick atmosphere could provide the same mass and radius than an ice-rich planet without atmosphere (Adams et al., 2008). The existence of a thick atmosphere can be explained either by an important outgassing during the planet history or by an erosion of a massive primordial atmosphere mostly composed of hydrogen and helium. The last case is very likely since a primordial atmosphere can be eroded by a strong stellar wind (Khodachenko et al., 2007). In the first case, the atmosphere might be composed mostly of $N_2$ and $CO_2$. But outgassing processes do not produce enough atmosphere to give a significant difference to the Earth-like case. According to solar-system examples (Venus, Earth, Titan), outgassing is limited in the sense that even for larger planets where more silicates are available, the amount of volatiles which can be outgassed through geological processes is very limited.

For transiting planets, both mass and radius can be determined. Mass measurement is provided by radial velocity and, in the near future, by astrometry. The precision on the planet mass derived from radial velocity measurements is dominated by the precision on the amplitude *K* of the radial velocity measurement. In the case of the 4.7 Earth mass planet HD 40307 b, it has been shown that a typical 5 % accuracy can be achieved with a 3.6 m telescope (Mayor et al. 2008). With high accuracy radial velocity spectrograph projects such as Codex at the E-ELT (Pasquini et al., 2005) it will be significantly better. The radius is inferred from the relative decrease of the stellar flux during the transit. The precision of the planet radius depends on the precision of the transit depth, itself depending on the photon noise for stable stellar flux, and on the precision of the parent star radius. The latter is the dominant factor as illustrated by the case of HD 189733 b: the precision of the stellar radius is 1.5 % while the precision on the transit depth (measured by the Hubble Space Telescope) is 0.27 % (Pont et al. 2007). We infer that for a planet with 2 Earth radius (1/5 $R_{Jup}$), the precision on the depth is 5 * 0.27 % = 1.35 %, leading to a radius estimate with $\sqrt{1.35^2 + 1.5^2}$ = 2 % accuracy. In summary, mass and radius of transiting planets can be determined with an accuracy of roughly 5 % and less than 3 %, respectively. But only 0.5 % of planets at 1 UA from a Solar type star make such transits. In these very specific cases, the classification of the planet will be straightforward.

For non transiting planets, only the mass *M* (resp. the product *M* sin *i*) can be estimated from astrometry (resp. radial velocity) observations. This information cannot provide any constraint on the nature of a Super-Earth (Mercury-like, Earth-like, water-rich, or mini-Neptune) as long as the radius is not estimated. This measurement will probably be obtained from « mutual events » for planets having a sufficiently large companion (Cabrera and Schneider, 2007), a situation which is very frequent in the solar system. But if mutual events are expected to occur in some cases, on must keep in mind that direct measurements of planetary radii will not be available in the general case. In the mid-term the most significant progress in exoplanetology will most likely come from the physicochemical characterization of exoplanets using spectro-polarimetric observations. It will then be possible to get two independent observations for each planet: the planetary mass (by radial velocity), and the flux. To obtain radius estimate from the flux will be difficult because many parameters are involved in the flux-radius relationships. The radius could be computed from the thermal flux through the relation:

$$F_{th}(\lambda) = 4\pi\sigma R^2 T_s^4 \, a(\lambda) \tag{5}$$

with $\sigma$ the Boltzmann constant, $T_s$ the surface temperature and $a(\lambda)$ an absorption factor reflecting the effective absorption by molecular species at wavelength $\lambda$. This expression does not take into account the possibility that the planet is surrounded by rings or possesses a companion, two factors which significantly contributes to the flux (e.g. Schneider, 2003). Radius could also be inferred from the reflected flux, but it relies on an estimate of the albedo:

$$F_{refl}(\lambda, P) = F_*(\lambda) \left(\frac{R}{a}\right)^2 \frac{A(\lambda, P)}{4} \Phi \tag{6}$$

where P is the state of polarization of the observed reflected light, Φ an orbital phase factor, and A a factor combining the planet reflectance and molecular absorption features. Here again this relation does not hold in case of rings and companions surrounding the planet (Arnold and Schneider, 2004; Cabrera and Schneider, 2007). Equations (5) and (6) illustrate that the thermal and reflected flux are proportional to the planet radius to the square. It is out of the goal of this work to discuss the real and complex modelling of these fluxes. The point is that in the general case, one can expect to get a rough estimate of the planet radius from the flux measurement, but most certainly with large uncertainties because of the many parameters involved in the relationships between the flux and the radius.

In the following discussion, we assume that the contribution of the atmosphere can be removed. In fact the detectability of the spectral signatures observable in transmission, depends on the molecular weight of the main atmospheric component, the planetary gravity field and the atmospheric temperature. As the temperature can be, in first approximation, derived from orbital characteristics, we are certainly able to distinguish a rocky/icy planet wrapped into an atmosphere from a gaseous one. Moreover, using the secondary transit technique in the visible spectral range, albedo can be measured (Rowe et al., 2006). If the mass of the planet is known and if the reflected flux can be measured together with the planetary albedo, the radius can be inferred by combining equations (3) and (5-6). If the radius estimate derived from the M-R relation differs significantly from the radius measured with the planet flux (thermal or reflected), it can only be due to the presence of a thick atmosphere. In the last case, M-R relationships allow the estimation of the size of the solid core by subtracting to the measured radius the theoretical radius derived from equation (3).

If for a planet from the Super-Earth family, both mass and radius are measured, and if the atmospheric contribution can be removed, then its nature is deduced from internal structure modelling.

As an example, the following discussion illustrates the degree of precision that can be achieved for the estimate of the water amount. The standard deviation of this estimate $\sigma_x$ is related to the standard deviation of the model $\sigma_x^{mod}$, and the precision of the M-R measurements : $(\sigma_x)^2 = (\sigma_x^{mod})^2 + (\sigma_x^{meas})^2$. In this relation, the second term has been computed in section III.4. The last term is estimated iteratively from equations (3-4) for any planetary mass. In the ideal case where both M and R have been measured exactly, the amount of water in the solid (or liquid) part of a Super-Earth will be known with a precision better than 4.5 % (figure 8; lower left corner of figure 10). In the general case, radius estimate of the solid part of the planet will be poorly constrained (especially if it is deduced from the removal of a thick atmosphere). But it will still be possible to get a rough estimate of the amount of water within a planet. Results are plotted in figure 10 in a $\sigma_R/R$ versus $\sigma_M/M$ diagram. It appears that radius estimates must be precise for having a good description of the planetary composition but that mass estimates can be less accurate, especially if the radius measurement is inaccurate. This is illustrated by the important flattening of the curves. For example, it can be seen that for a relative radius uncertainty of 3 %, the amount of water can be determined with a precision between 10 % and 12.5 % depending on the mass uncertainty. These results, in good agreement with the previous works from Seager et al. (2007) and Valencia et al. (2007b), clearly illustrate the importance of having good radius measurements for characterization purposes. It is only below 2% of relative accuracy on radius measurement (which is probably not achievable in near future except for transiting planets), that the accuracy of mass measurement will provide more constraints on the composition of the planet.

## V. Conclusion

In this paper, internal structure modelling of planetary interiors has been used to investigate the mass – radius relations for Super-Earths planets. A general law has been proposed which describes the mass-radius relation with an accuracy of 1 % in a nominal case from 1 to 100 Earth masses and for any amount of water ice in the planet. Numerical results are in good agreements with previous works and fill the link between the studies devoted to planets with very low masses and those developed for very large masses.

In a second part, the uncertainties on radius estimates due to model approximation have been investigated. Compositional variations in silicates and in iron alloys change the relative size of the metallic core but do not influence strongly the planetary radius (less than 2 %). Similarly, very large thermal variations within the planetary layers do not affect significantly the size of the planet (1.5 %). A combination of these two factors indicates that numerical modelling provides radii estimate of dry silicate – rich planets with a precision better than 5 %. Since it will be very hard to provide any constraints on the silicate composition and on the thermal profiles, one cannot expect to have a more precise constraint from numerical modelling.

The M-R relation proposed in this work will allow the characterization of the main composition of the solid cores of exoplanets as long as both mass and radii can be measured, and if the contribution of the atmosphere can be removed. This drastic conditions are not yet fulfilled but the different approaches that will allow such determinations in the future have been discussed in this work. In addition, a diagram showing the standard deviation on the water amount estimate as a function of planetary mass and radius uncertainties has been proposed.

## *VI. Acknowledgments*

The authors want to thank G. Tinetti for her valuable comments on detection perspectives and an anonymous reviewer for her (his) constructive remarks, Part of this work was carried out at the Jet Propulsion Laboratory, California Institute of Technology, under contract with NASA.

**References :**

Adams , Seager S., Elkins-Tantor. 2008, *Astrophys. J.*, 673, (2), 1160

Allègre C.J., Poirier J.P., Humler E., Hofmann A.W. 1995, 134: 515.

Arnold L. and Schneider J. 2004, Astron. Astroph., 420, 1153.

Baraffe, I.; Chabrier, G.; Barman, T. 2008, Astron. & Astrophys., 482 (1), 315.

Beaulieu, J.-P.; Bennett, D. P.; Fouqué, P.; Williams, A.; Dominik, M.; Jørgensen, U. G.; Kubas, D.; Cassan, A.; Coutures, C.; Greenhill, J.; Hill, K.; Menzies, J.; Sackett, P. D.; Albrow, M.; Brillant, S.; Caldwell, J. A. R.; Calitz, J. J.; Cook, K. H.; Corrales, E.; Desort, M.; Dieters, S.; Dominis, D.; Donatowicz, J.; Hoffman, M.; Kane, S.; Marquette, J.-B.; Martin, R.; Meintjes, P.; Pollard, K.; Sahu, K.; Vinter, C.; Wambsganss, J.; Woller, K.; Horne, K.; Steele, I.; Bramich, D. M.; Burgdorf, M.; Snodgrass, C.; Bode, M.; Udalski, A.; Szymański, M. K.; Kubiak, M.; Wi ckowski, T.; Pietrzyński, G.; Soszyński, I.; Szewczyk, O.; Wyrzykowski, Ł.; Paczyński, B.; Abe, F.; Bond, I. A.; Britton, T. R.; Gilmore, A. C.; Hearnshaw, J. B.; Itow, Y.; Kamiya, K.; Kilmartin, P. M.; Korpela, A. V.; Masuda, K.; Matsubara, Y.; Motomura, M.; Muraki, Y.; Nakamura, S.; Okada, C.; Ohnishi, K.; Rattenbury, N. J.; Sako, T.; Sato, S.; Sasaki, M.; Sekiguchi, T.; Sullivan, D. J.; Tristram, P. J.; Yock, P. C. M.; Yoshioka, T. 2006, Nature, 439 (7075), 437.

Beirao P., Santos N.C., Isralian G., Mayor M. 2005, Astron. & Astrophys., 438, 251.

Belonoshko A. and Saxena S.K. 1991, *Geochem. Cosmoch. Acta*, 381.

Bennett, D. P.; Bond, I. A.; Udalski, A.; Sumi, T.; Abe, F.; Fukui, A.; Furusawa, K.; Hearnshaw, J. B.; Holderness, S.; Itow, Y.; Kamiya, K.; Korpela, A. V.; Kilmartin, P. M.; Lin, W.; Ling, C. H.; Masuda, K.; Matsubara, Y.; Miyake, N.; Muraki, Y.; Nagaya, M.; Okumura, T.; Ohnishi, K.; Perrott, Y. C.; Rattenbury, N. J.; Sako, T.; Saito, To.; Sato, S.; Skuljan, L.; Sullivan, D. J.; Sweatman, W. L.; Tristram, P. J.; Yock, P. C. M.; Kubiak, M.; Szymanski, M. K.; Pietrzynski, G.; Soszynski, I.; Szewczyk, O.; Wyrzykowski, L.; Ulaczyk, K.; Batista, V.; Beaulieu, J. P.; Brillant, S.; Cassan, A.; Fouque, P.; Kervella, P.; Kubas, D.; Marquette, J. 2008, *Astrophys. J.*, 2008arXiv0806.0025B.

Benz W. 2006, Meteoritics & Planetary Science, 41, Supplement, Proceedings of 69th Annual Meeting of the Meteoritical Society, 5393.

Birch F. 1964, *J. Geophys. Res*. 69: 4377.

Cabrera J. and Schneider J. 2007, Astron. & Astrophys., 464, 1133.

Cox et al., 1999, Allen's astrophysical quantities, 4th edition, Arthur N. Cox Editor, Springer – Verlag.

Dreibus G., Palme H. 1996, Geochimica et Cosmochimica Acta, 60, (7), 1125.

Erkaev N., Lammer H., Kulikov Y., Langmayr D., Selsis F., Jaritz G. & Biemat H. 2007, Astron. & Astrophys., 472 , 329

Fortney J.J., Marley M.S., Barnes J.W. 2007, Astroph. J., Volume 659 (2), 1661.

Gilli G., Israelian G., Ecuvillon A., Santos N.C., Mayor M. 2006, *Astron. Astroph.*, 449 (2), 723.

Hemley R.J., Jephcoat A.P., Mao H.K., Zha C.S., Finger L.W., Cox D.E. 1987,. *Nature*, 330, 737.

Huang S. 1960, Publ. Astron. Soc. Pacific, 72, 489.

Javoy M. 1995, *Geophys. Res. Lett*., 22(16): 2219.

Karato S. 2004, American Geophysical Union, Fall Meeting 2004, abstract #T44B-01.

Khodachenko M., Ribas I., Lammer H., et al. 2007, Astrobiology, 29, 167.

Léger A., Selsis F., Sotin C., Guillot T., Despois D., Mawet D., Ollivier M., Labèque A., Valette C., Brachet F., Chazelas B., Lammer H. 2004, *Icarus,* 169, 499.

Lodders, K. 2003, *Astrophys. J.*, 591, 1220

Lovis, C., M.Mayor, Pepe, F., Alibert, Y., Benz, W., Bouchy, F., Correia, A. C. M., Laskar, J., Mordasini, C., Queloz, D., Santos, N. C., Udry, S., Bertaux, J., & Sivan, J. 2006, Nature, 441, 305.

Manning C.E. 2006, Geochimica et Cosmochimica Acta, 70 (18), 388.

Mayor, M. 2008, Super-Earths workshop, Nantes, June 2008.

Mayor M., Udry S., Lovis C., Pepe F., Queloz D., Benz W., Bertaux J.-L., Bouchy F., Mordasini C., & Segransan D., 2008. Astron. & Astrophys. submitted .

McArthur, Barbara E.; Endl, Michael; Cochran, William D.; Benedict, G. Fritz; Fischer, Debra A.; Marcy, Geoffrey W.; Butler, R. Paul; Naef, Dominique; Mayor, Michel; Queloz, Diedre; Udry, S.; Harrison, Thomas E. 2004, *Astrophys. J.*, 614 (1), 81.

Melnick et al. 2001, 199th AAS Meeting, #09.10

Murakami M., Hirose K., Kawamura K., Sata N., Ohishi Y. 2004, Science, 304, 855.

Musselwhite D. and Drake M. 2000,. Icarus, 148, 160

Ohtani E., Ringwood A.E., and Hibberson W. 1984, Earth Planet. Sci. Lett., 71: 94.

Pasquini L, Cristiani S., Dekker H., Haehnelt M., Molaro P., Pepe F., Avila G., Delabre B., D'Odorico V.,Vanzella, E. and 12 coauthors 2005. The ESO Messenger no 122 p. 10

Poirier J.-P.. 2000, Cambridge Univ. Press, 2$^{nd}$ ed., Cambridge, UK.

Pont F., Gilliland R., Moutou C., Charbonneau D., Bouchy F., Brown T., Mayor M., Queloz D., Santos N. & Udry S. 2007. Astron. & Astrophys., 467, 1347

Rivera, E. J.; Lissauer, J. J.; Butler, R. P.; Marcy, G. W.; Vogt, S. S.; Fischer, D. A.; Brown, T. M.; Laughlin, G.; Henry, G. W. 2005, Bull. Am. Astron. Soc., 37, 1487.

Rowe, Jason F.; Matthews, Jaymie M.; Seager, Sara; Kuschnig, Rainer; Guenther, David B.; Moffat, Anthony F. J.; Rucinski, Slavek M.; Sasselov, Dimitar; Walker, Gordon A. H.; Weiss, Werner W. 2006, *Astrophys. J.*, 646 (2), 1241.

Salpeter E.E., and Zapolsky H.S. 1967, *Physical Review II*, 158, 876.

Santos, N. C.; Bouchy, F.; Mayor, M.; Pepe, F.; Queloz, D.; Udry, S.; Lovis, C.; Bazot, M.; Benz, W.; Bertaux, J.-L.; Lo Curto, G.; Delfosse, X.; Mordasini, C.; Naef, D.; Sivan, J.-P.; Vauclair, S. 2004, Astron. & Astrophys., 426 (9), 19.

Schneider J. 1999, C. R. Acad. Sci. Paris 327, 621.

Schneider J. 2003. in Proceedings of Towards other Earths: Darwin/TPF and the search for extrasolar terrestrial planets , ESA SP-539, p. 205

Seager S., Kuchner M., Hier-Majumder C. A., Militzer B. 2007, *Astrophys. J.,* 669 (2), 1279.

Selsis, F.; Chazelas, B.; Bordé, P.; Ollivier, M.; Brachet, F.; Decaudin, M.; Bouchy, F.; Ehrenreich, D.; Grießmeier, J.-M.; Lammer, H.; Sotin, C.; Grasset, O.; Moutou, C.; Barge, P.; Deleuil, M.; Mawet, D.; Despois, D.; Kasting, J. F.; Léger, 2007, Icarus, 191 (2), 453.

Shim S.H. 2008, *Ann. Rev. Earth Planet. Sci*. (36), 569.

Sohl F., Spohn T. 1997, J. Geophys. Res., 102 (E1), 1613.

Sotin C., Grasset O., Mocquet A. 2007, *Icarus*, 191 (1), 337.

Tajika E. 1998, Geophys. Res. Lett.25(21), 3991.

Thompson S.L. 1990, *Sandia Natl. Lab. Doc.*, SAND89-2951.

Udry, S., Bonfils, X., Delfosse, X., Forveille, T., Mayor, M., Perrier, C., Bouchy, F., Lovis, C., Pepe, F., Queloz, D., & Bertaux, J.-L. 2007, Astron. & Astrophys., 469 (3), L43.

Valencia D., O'Connel R.J., Sasselov D. 2006, *Icarus*, 181, 545.

Valencia D., Sasselov D., O'Connell R.J. 2007a, Astrophys. J. 656, 545.

Valencia D., Sasselov D.; O'Connell R., Astrophys. J., 2007b, 665 (2), 1413-1420.

Zapolsky H.S., Salpeter E.E. 1969, *Astrophys. J.*, 158, 809.

Zuckerman B., Forveille T. & Kastner J. 1995, Nature, 373, 494.

**Figure captions:**

**Figure 1**: Internal structure of planets: A fully differentiated planet is composed of a metallic core, surrounded by a silicate mantle and possibly an icy mantle. Both silicates and water ices can be split into two sublayers corresponding to low and high pressure phases. The low pressure silicate layer influences the global size of the planet if the amount of water is small and the planetary mass is below 10 $M_E$. For a large amount of water, the pressure at the base of the icy mantle is above the pressure transition from olivine to perovskite.

**Figure 2**: Distribution of Fe/Si and Mg/Si for stars which hosts exoplanets (black diamonds). Measured ratio are corrected by adding the contribution of Ca, Al, and Ni to their closest major element. The distribution is located around Fe/Si=1.01 and Mg/Si = 1.34 (white dot). The solar value is indicated by the white square. Three domains have also been plotted depending on the size of the core (1 %, 20 %, and 40 % respectively). On each domain, variations of the $Mg_\#$ parameter are indicated (small numbers on dashed lines). It appears that the petrological model used for describing both metallic core and silicate mantles allows to investigate the whole range of composition of stellar systems which are known to host exoplanets.

**Figure 3**: Density variation at 300 K as a function of pressure for pure iron (a), forsterite $MgSiO_3$ (b), and pure water (c). TFD curves are from Seager et al. (2007). ANEOS curves are obtained using the ANEOS package (Thompson, 1990). Mïe-Gruneisen-Debye estimate are derived from Sotin et al. (2007). In addition, data from Belonoshko and Saxena (1991) have been added in the supercritical domain for the water compound (dashed line). For each material, the grey domain indicates the plausible pressure range relevant to planetary interiors below 100 $M_E$.

**Figure 4**: M-R relationships below 100 $M_E$ in the reference case. Numbers indicate the amount of water in %wt ($X_w$). Black diamonds: Model results from equation (2). Black lines: M-R relationships using equations (3-4) with parameters from Table 2. Uranus, Neptune and GJ436 B are indicated by white dots.

**Figure 5**: Effect of composition on radii in the case of 10 $M_E$. These curves are almost insensitive to planetary mass. $R_0$ and $R_{c0}$ are the planetary and the metallic core radii respectively in the reference case (eq. 3-4) for a dry planet ($X_w$=0). For 10 $M_E$, $R_0$=12321 km and $R_{c0}$= 5483 km. The radius of the planet is almost insensitive to the amount of volatiles in the core and the $Mg_\#$, but varies slightly (± 2%) with (Mg/Si) and (Fe/Si). Core radius depends strongly on the compositional parameters, and especially on $Mg_\#$.

**Figure 6**: *Top*: The investigated thermal domain. Ninety thermal profiles have been generated randomly within the grey area by changing the adiabatic slope in each layer and the temperature drop through the interfaces. *Bottom*: Core and planetary radii for each thermal case assuming a 10 Earth mass planet. It shows that thermal constraints are of second order importance for describing M-R relation of massive exoplanets. Results do not depend on the mass of the planet. The reference radii $R_0$ and $R_{c0}$ are defined as in figure 5.

**Figure 7**: *Top*: M-R distribution of the 530 simulated Earth-like planets. The points are randomly distributed within the M-R domain described by equations (3-4). *Bottom*: Compositional variations of the simulated planets as a function of their mass. Compositions are fixed by the four independent parameters $Mg_\#$, Fe/Si, Mg/Si and $X_w$, which are allowed to vary within the ranges defined in Table 1. In addition, temperature profiles are varied randomly for each planet with respect to the standard deviations proposed in table 1 for both adiabatic slopes and temperature drops through the interfaces (not plotted).

**Figure 8**: *Top*: Comparison between the amount of water $X_w$ imposed in the model and the computed value $X_w^*$ assuming that the planet is in the reference case. Data are obtained using the 530 simulated cases plotted in figure 7. *Bottom*: Gaussian distribution of the $X_w$-$X_w^*$ differences for $\sigma_x^{\text{mod}} = 0.045$. It illustrates the fact that, if radius and mass of a Super-Earth are exactly known, its amount of water can be estimated with an accuracy of 4.5 %.

**Figure 9**: Mass-radius relationships for solid planets composed of iron, silicates, and ices (shaded areas). The amount of water varies from 0 % to 100 % from bottom to top of the white domain limited by the silicates curves at the bottom and the pure ices curve at the top (see figure 4). Results from previous studies and for several pure compounds (H, He, C, Fe, Mg) are also plotted for a purpose of comparison. Position of GJ436b and solar gaseous planets are indicated by white circles.

**Figure 10**: Accuracy of the water amount measurement using M-R relationships in the case of Super-Earths . Small numbers indicate the value of the standard deviation $\sigma_{Xw}/X_w$ depending on the precision on mass and radius measurements. The bottom left corner, (at 4.5 %), corresponds to the standard deviation related to the internal structure modelling (choice of thermal profiles, equation of states, compositions of silicate and iron alloy phases).

| **Planet description** | **Reference Case** | **Allowed ranges** |
|---|---|---|
| *Global composition* | | |
|     Fe/Si | 1.01 | ± 0.4 |
|     Mg/Si | 1.34 | ± 0.6 |
|     %wt of water ($X_w$) | 0.1 - 80 | |
| *Metallic core* | | |
|     % volatiles | none | unchanged |
|     EOS | ANEOS | unchanged |
|     Upper Temp. drop | 800 K | ± 500 K |
|     Temperature profile | adiabatic | ± 2 % |
| *Lower silicate mantle* | | |
|     $Mg_\#$ | 0.8 | ± 0.2 |
|     EOS | ANEOS above 100 GPa<br>MGD below 100 GPa | unchanged |
|     Upper Temperature drop | 300 K | ± 200 K |
|     Temperature profile | adiabatic | ± 2 % |
| *Upper silicate mantle* | | |
|     $Mg_\#$ | 0.8 | ± 0.2 |
|     EOS | 3$^{rd}$ order Birch-Mürnaghan | unchanged |
|     Upper Temperature drop | 1300 K | ± 500 K |
|     Temperature profile | adiabatic | unchanged |
| *Icy mantle* | | |
|     % volatiles | none | unchanged |
|     EOS | MGD+ ANEOS+ transitional regime | unchanged |
|     Upper Temp. drop | 100 K | ± 100 K |
|     Temperature profile | adiabatic | ± 2 % |
| *Liquid water* | | |
|     % volatiles | none | unchanged |
|     EOS | 3$^{rd}$ order Birch-Mürnaghan | unchanged |
|     Surface temperature | 300 K | ± 200 K |
|     Temperature profile | adiabatic | ± 2 % |

**Table 1**: Input parameters and main characteristics of the reference case. It corresponds to a planet which possesses a global composition equal to the averaged ratio of the stars studied by Beirao et al. (2005) and Gilli et al. (2006). It is an "extreme case" because the metallic core is composed of pure iron (no volatiles). The third column indicates the ranges in which the different parameters have been randomly distributed for studying the accuracy of the model in section III.

|  | α | β | γ | ε |
|---|---|---|---|---|
| $\xi_2$ | -1.704 $10^{-5}$ | -9.015 $10^{-7}$ | 5.397 $10^{-8}$ | -3.166 $10^{-10}$ |
| $\xi_1$ | 5.714 $10^{-3}$ | -2.116 $10^{-4}$ | -4.221 $10^{-6}$ | 3.085 $10^{-8}$ |
| $\xi_0$ | 1.010 | 2.859 $10^{-1}$ | -5.518 $10^{-4}$ | 2.096 $10^{-6}$ |

**Table 2**: Parameters used in equation 4. These values provide a description of M-R relations in the mass range [1-100 $M_E$] for any amount of ices (from the Earth-like case to the pure water world).

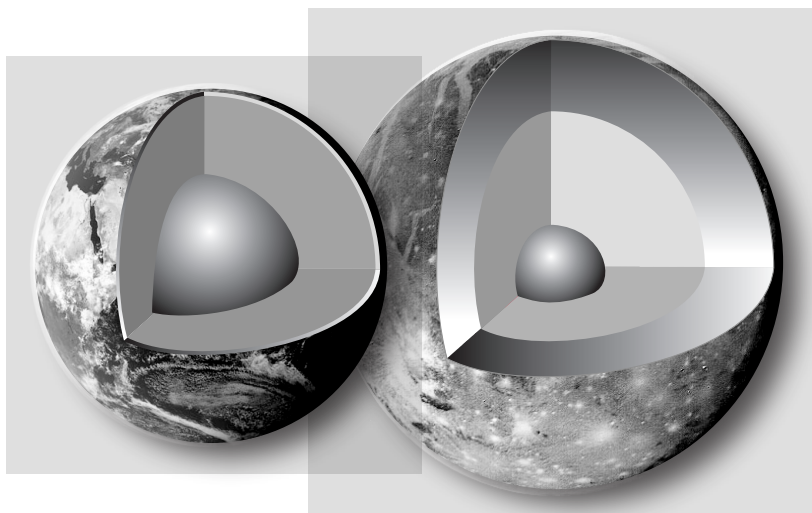

Figure 1

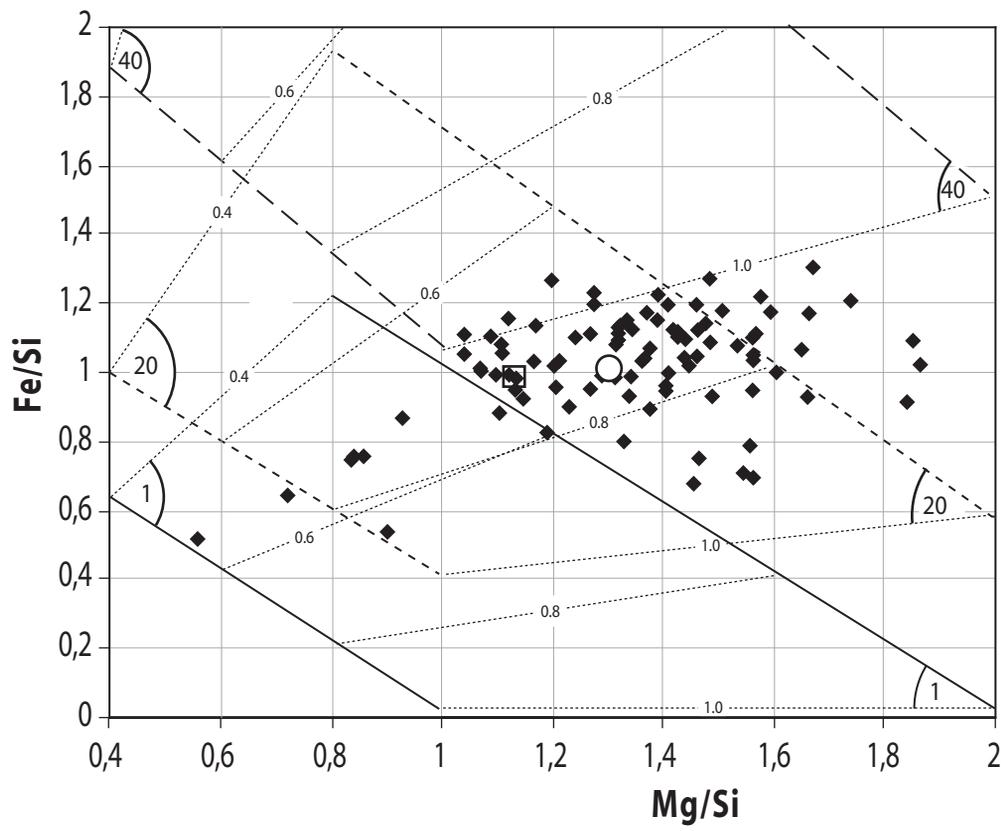

Figure 2

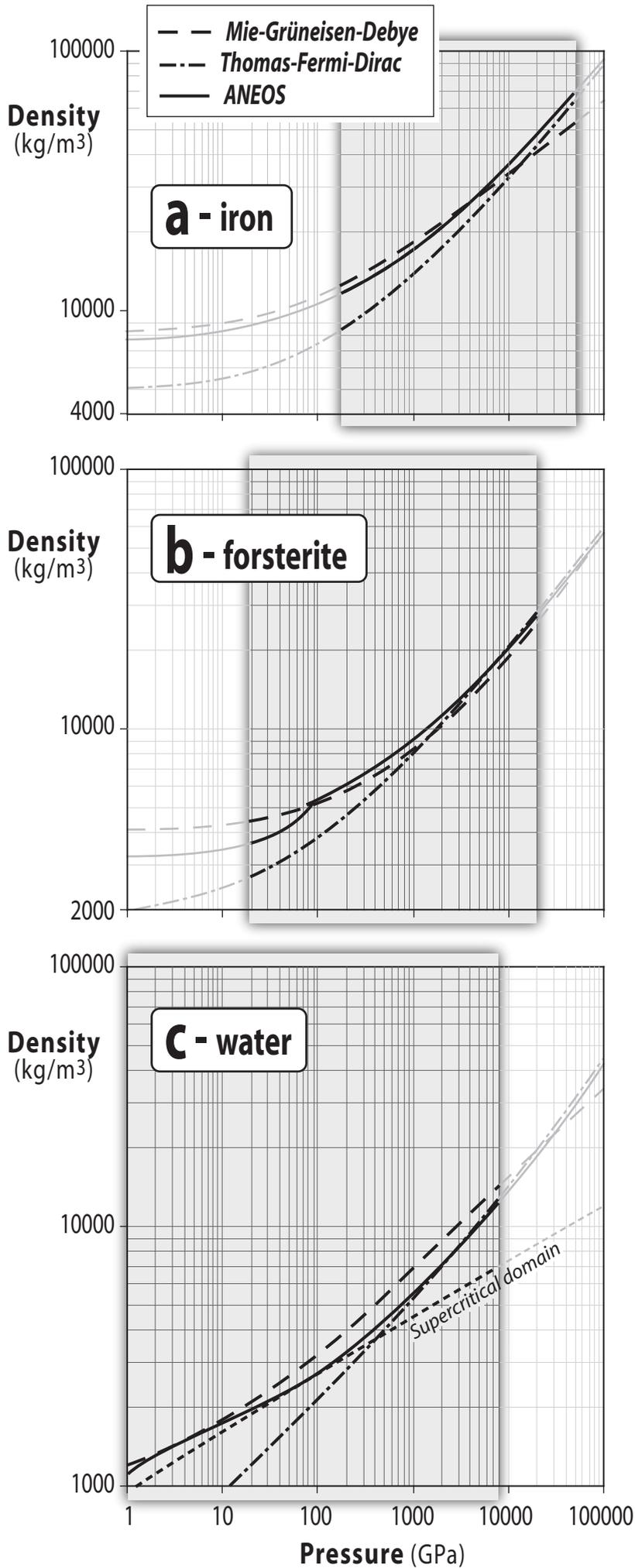

Figure 3

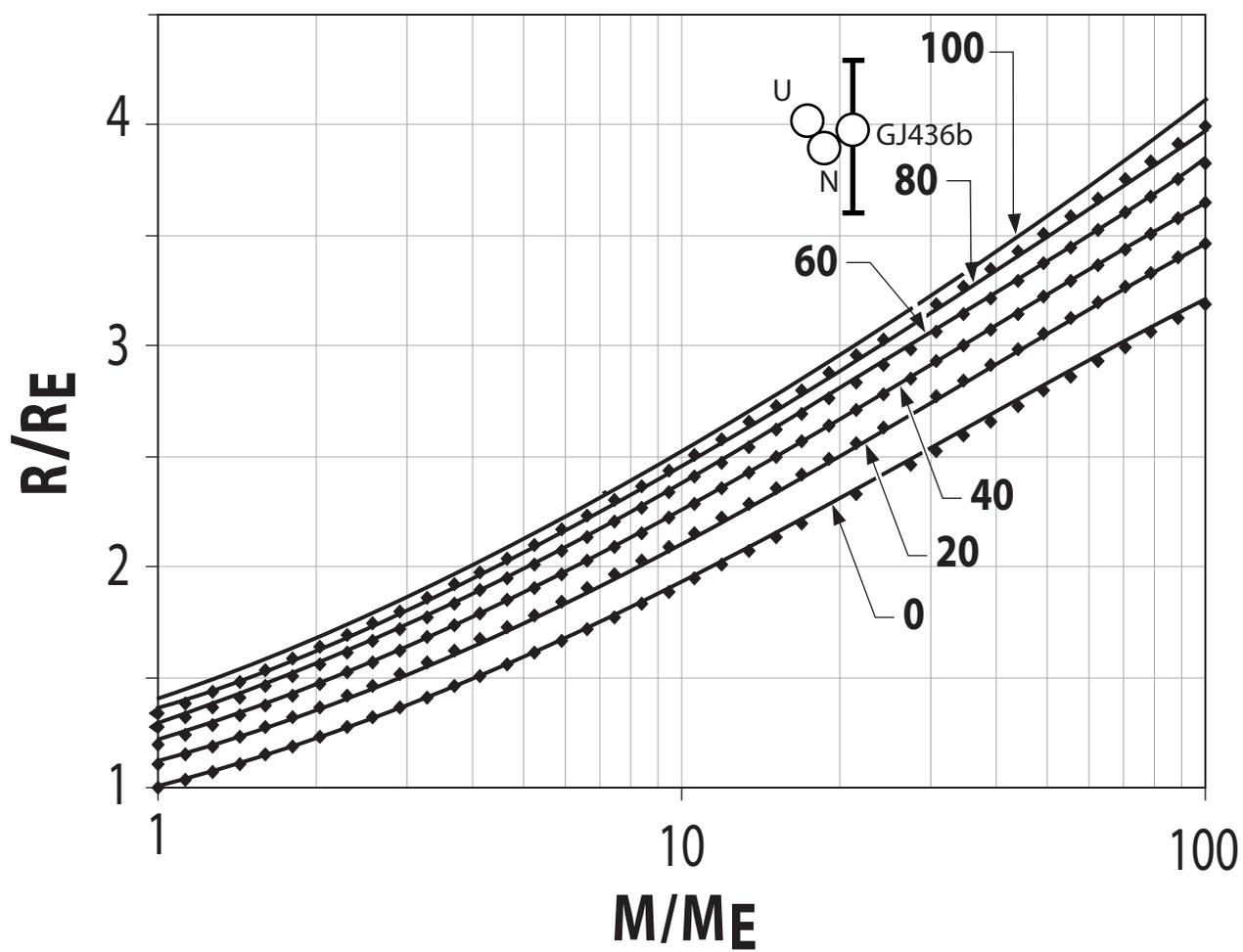

Figure 4

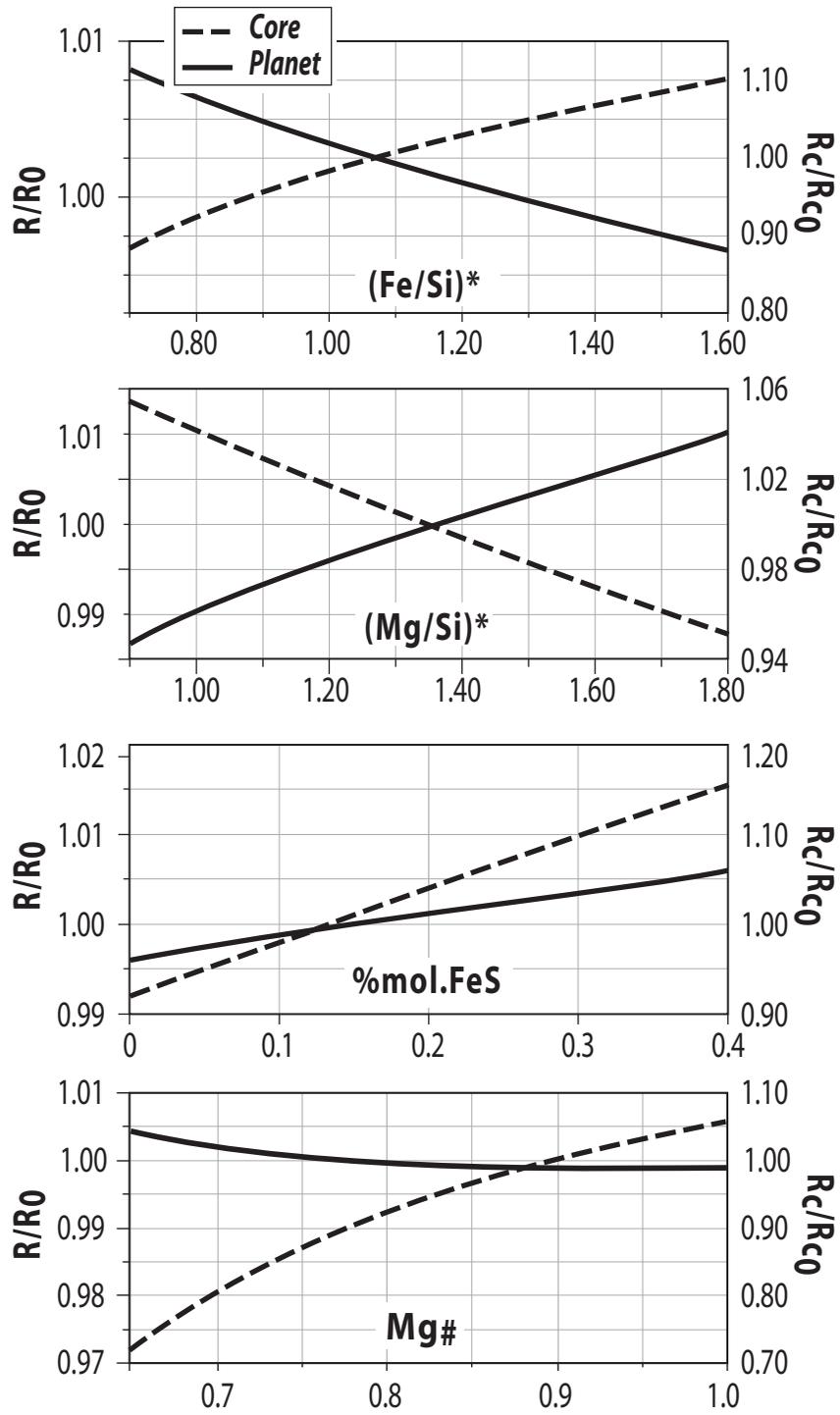

Figure 5

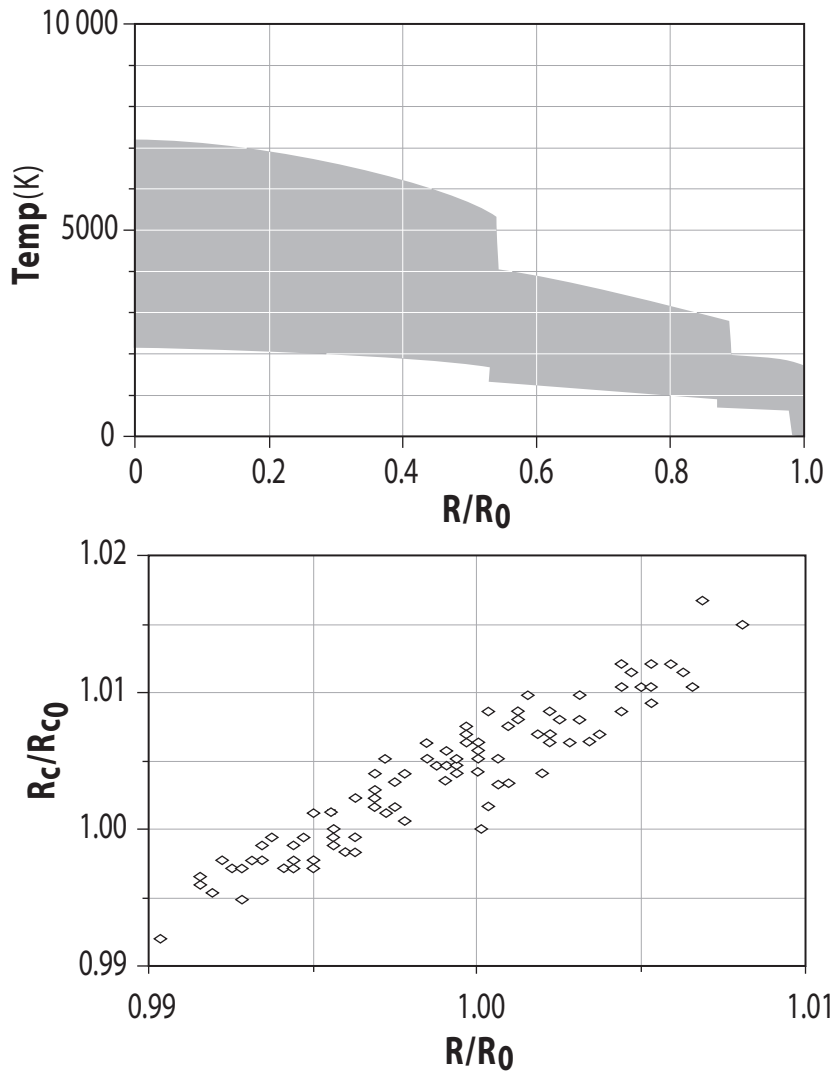

Figure 6

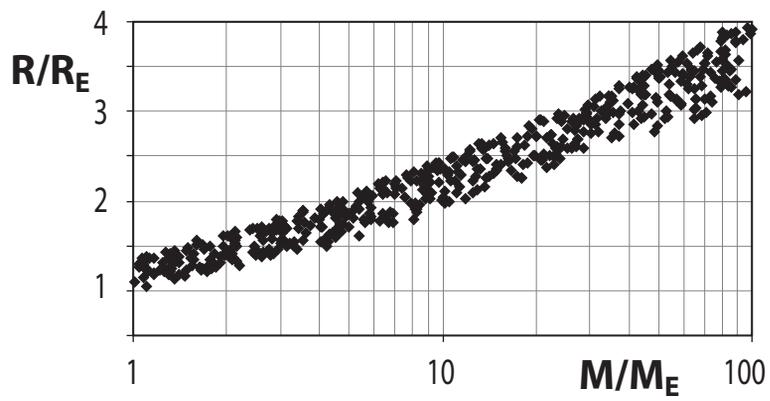
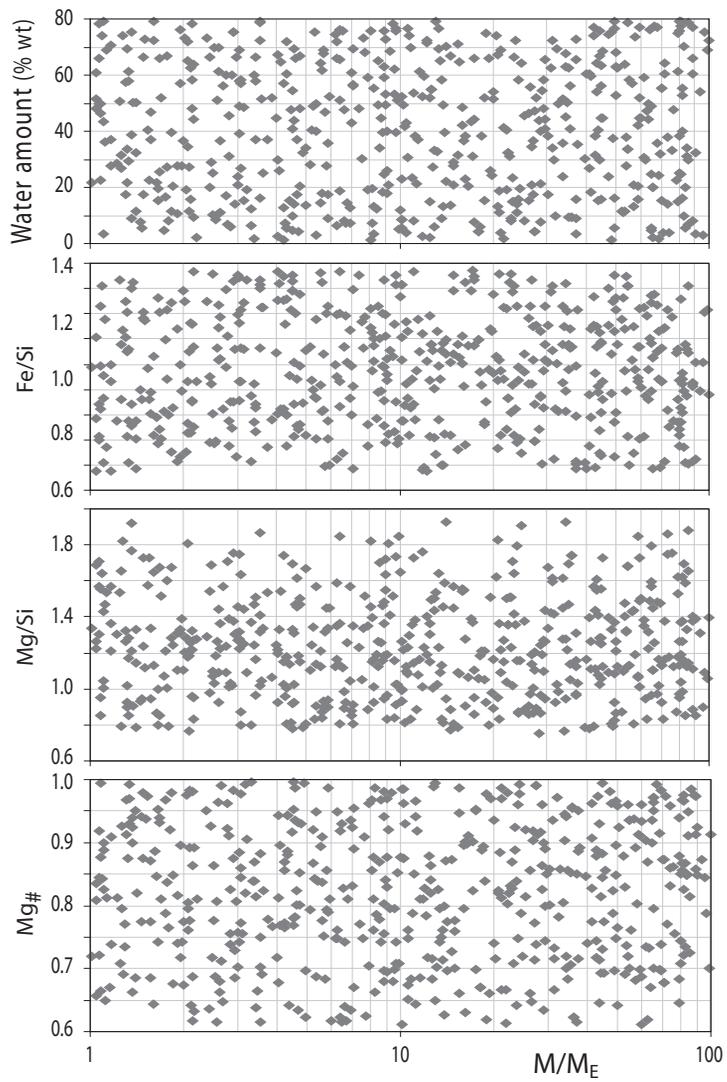

Figure 7: Grasset et al.

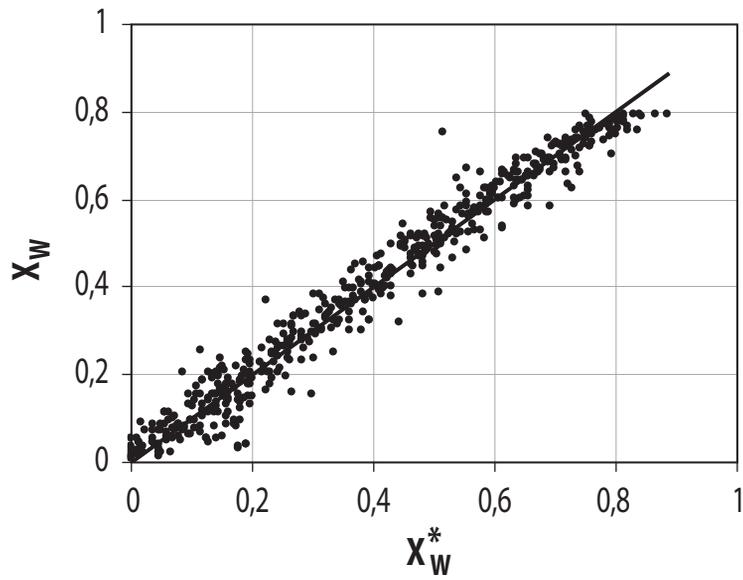
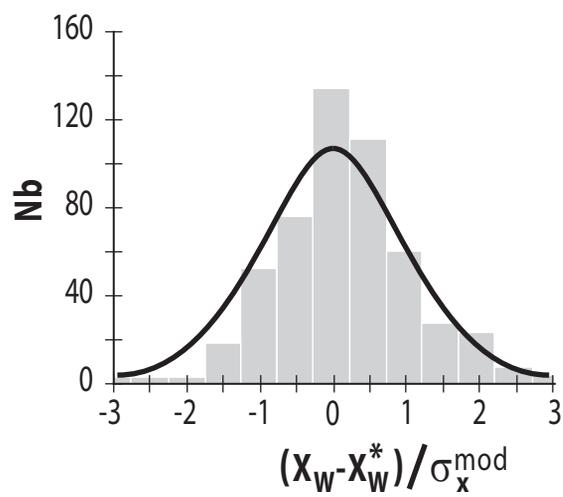

Figure 8: Grasset et al.

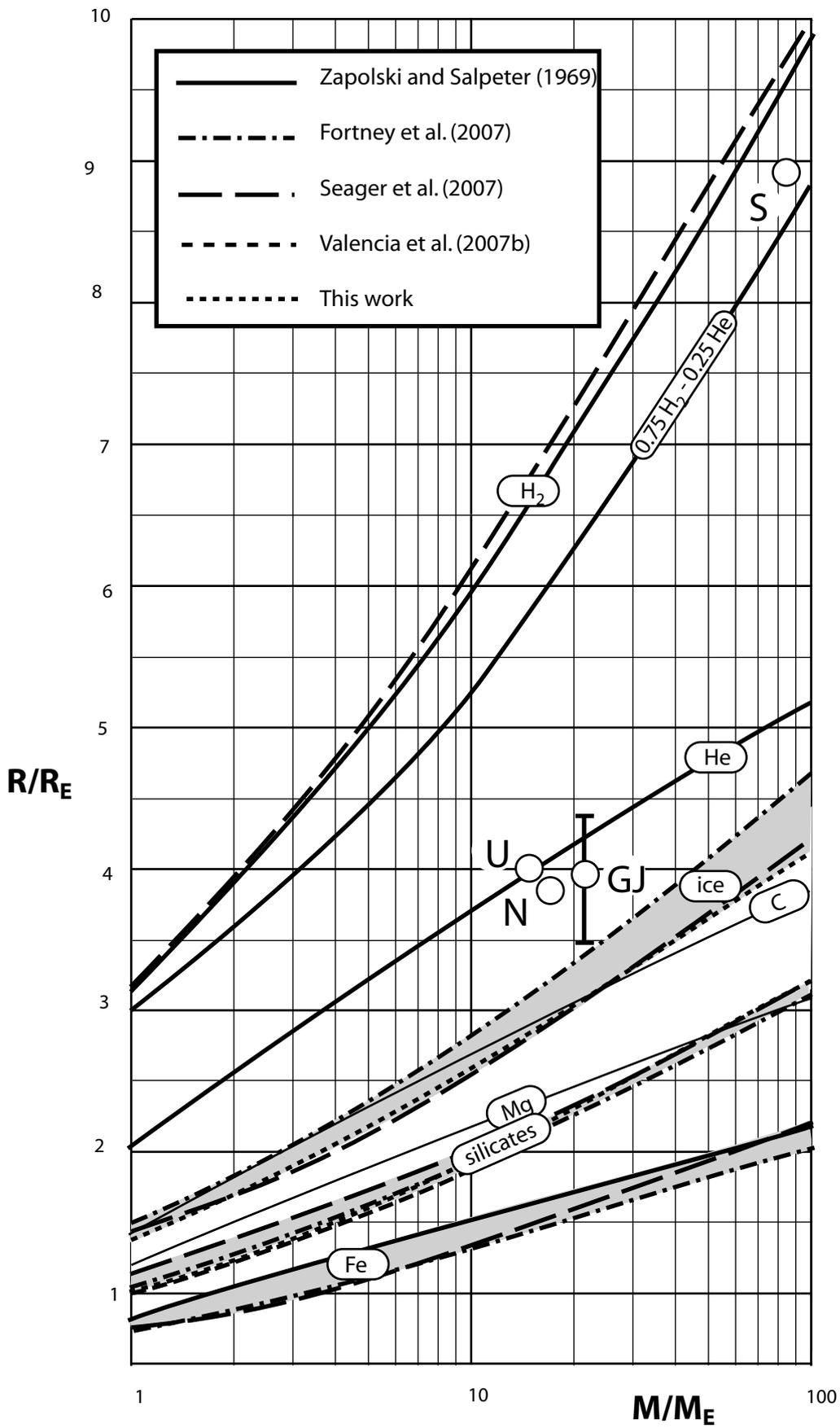

Figure 9

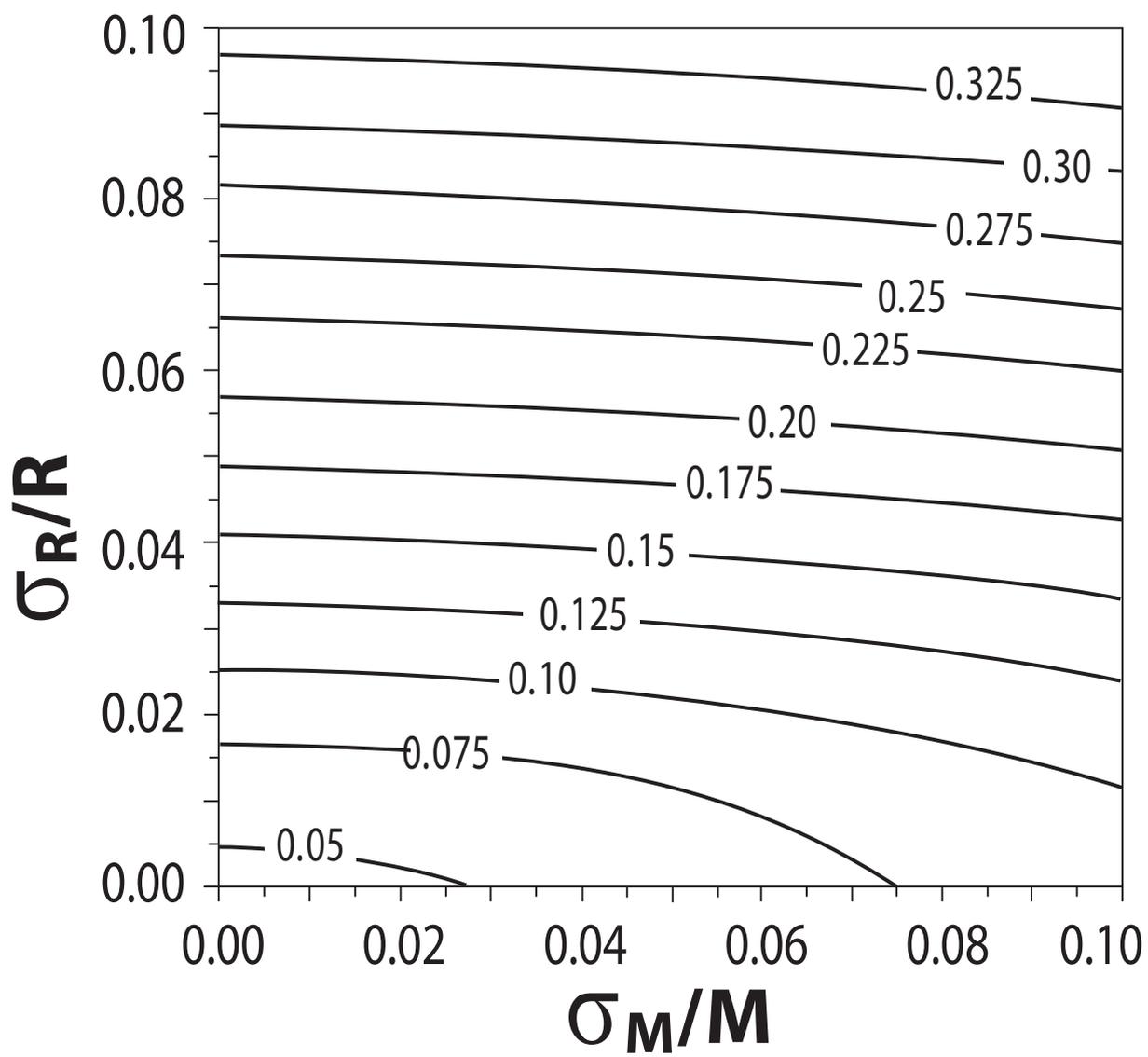